%% file: main.tex
\documentclass[AMA,LATO1COL]{WileyNJD-v2}

\articletype{Article Type}%

\received{26 April 2016}
\revised{6 June 2016}
\accepted{6 June 2016}

\usepackage{csquotes}
\usepackage{listings}
\usepackage{color}
\usepackage{todonotes}

\definecolor{dkgreen}{rgb}{0,0.6,0}
\definecolor{gray}{rgb}{0.5,0.5,0.5}
\definecolor{mauve}{rgb}{0.58,0,0.82}

\begin{document}

\title{Further Investigation of the Survivability of Code Technical Debt Items
}

\author[1]{Ehsan Zabardast*}

\author[2]{Kwabena Ebo Bennin}

\author[1]{Javier Gonzalez Huerta}

\authormark{Zabardast \textsc{et al.}}

\address[1]{\orgdiv{Software Engineering Research Lab (SERL)}, \orgname{Blekinge Institute of Technology}, \orgaddress{\state{Blekinge}, \country{Sweden}}}

\address[2]{\orgdiv{Information Technology Group}, \orgname{Wageningen University and Research}, \orgaddress{\state{Wageningen}, \country{The Netherlands}}}

\corres{*Ehsan Zabardast, Software Engineering Research Lab (SERL), Blekinge Institute of Technology, Campus Gr{\"a}svik, Valhallav{\"a}gen 1, Karlskrona, Sweden . \email{ehsan.zabardast@bth.se}}

\abstract[Summary]{
Context: Technical Debt (TD) discusses the negative impact of sub-optimal decisions to cope with the need-for-speed in software development. Code Technical Debt Items (TDI) are atomic elements of TD that can be observed in code artefacts. Empirical results on open-source systems demonstrated how code-smells, which are just one type of TDIs, are introduced and ``survive'' during release cycles. However, little is known about whether the results on the survivability of code-smells hold for other types of code TDIs (i.e., bugs and vulnerabilities) and in industrial settings.

Goal: Understanding the survivability of code TDIs by conducting an empirical study analysing two industrial cases and 31 open-source systems from Apache Foundation. 

Method: We analysed 133,670 code TDIs (35,703 from the industrial systems) detected by SonarQube (in 193,196 commits) to assess their survivability using survivability models.

Results: In general, code TDIs tend to remain and linger for long periods in open-source systems, whereas they are removed faster in industrial systems. Code TDIs that survive over a certain threshold tend to remain much longer, which confirms previous results. Our results also suggest that bugs tend to be removed faster, while code smells and vulnerabilities tend to survive longer.
}



\jnlcitation{\cname{%
\author{Zabardast E.}, 
\author{K.E. Bennin}, and
\author{J. Gonzalez Huerta}} (\cyear{2021}), 
\ctitle{Further Investigation of the Survivability of Code Technical Debt Items}, \cjournal{JSME}, \cvol{2021;00:1--6}.}

\maketitle


\input{sections/01_introduction}
\input{sections/02_relatedwork}
\input{sections/03_research_methodology}
\input{sections/04_results}

\input{sections/05_discussion}
\input{sections/06_theats}
\input{sections/07_conclusions}


\section*{Acknowledgements}
This research was supported by the KK foundation through the SHADE KK-H\"{o}g project under grant 2017/0176 and Research Profile project SERT under grant 2018/010  at Blekinge Institute of Technology, SERL Sweden.





\bibliography{references}%



\end{document}

%% file: sections/01_introduction.tex
\section{Introduction}\label{sec:introduction}

There is an ever-increasing pace in the size and complexity of software systems as they evolve, as Lehman~\cite{Lehman1979,Lehman1996} formulated in his Laws of Software Evolution, also known as Lehman's Laws. As a consequence of this ever-increasing size and complexity, software systems accumulate Technical Debt as they are developed and evolve~\cite{Kruchten2019}. Technical Debt (TD)~\cite{cunningham1993wycash} is a metaphor commonly used to discuss the negative impact of sub-optimal design decisions, often taken to cope with the need for speed in the development. As a software system evolves, these sub-optimal design decisions taken in order to be able to deliver the product in time can potentially hinder its maintainability and even our ability to deliver future releases of the product~\cite{Kruchten2019}.

Sub-optimal design decisions are often not visible; however, they might manifest in the form of Technical Debt Items (TDIs), which are the manifestations of TD~\cite{Kruchten2019}. TDIs can take the form of vulnerabilities and code smells~\cite{brown2010managing,Kruchten2012,lim2012balancing}, whilst some of them might be visible, and they might materialise in terms of bugs or defects~\cite{Kruchten2012}. TD items are ``atomic elements of TD'' that connect a set of artefacts (e.g., code) with the consequences of the quality~\cite{Kruchten2019} and have been introduced as a way to quantify or visualise TD. Bugs, code smells, and vulnerabilities are some examples of code Technical Debt Items (TDIs)~\cite{Kruchten2019,Li2015,saarimaki2019diffuseness}. These code TDIs go a long way to affect the development process and evolution of the software system, thus creating friction~\cite{Kruchten2019} and increasing the maintenance effort~\cite{sjoberg2012quantifying}. Adding new functionality to the existing system is also challenging when code TDIs are present in the system~\cite{neamtiu2013towards}.

Several empirical studies acknowledge the negative effects of code TDIs on software quality, e.g., \cite{chatzigeorgiou2010investigating,menzies2006data}. The repaid code TD (removing code TDI) not only improves the maintainability of the software system but also reduces the maintenance costs~\cite{sjoberg2012quantifying}. Studying the life cycle of code TDIs (bugs, code smells, and vulnerabilities) can therefore help prioritise the maintenance activities~\cite{tufano2017and}. Understanding the life cycle of code TDIs is essential in building support tools as well. The empirical analysis of the evolution and survivability (i.e., the time that the code TDIs remain in the system) of the code TDIs thus have crucial implications for software development teams.

Prior research~\cite{arcoverde2011understanding,brown2010managing,chatzigeorgiou2010investigating,digkas2018developers,lozano2007assessing,rapu2004using,tufano2015and,tufano2017and} have studied the survivability of code TDIs in software systems - mainly focusing on -- some -- code smells. While studies are investigating the effects of code TDIs on codebase and code entities, few empirical studies have been conducted on industrial settings. Additionally, a more in-depth investigation and insights into how development and maintenance activities impact the survivability of code TDIs are yet to be considered by prior studies. By taking inspiration and following a similar approach of the study by Tufano et~al.~\cite{tufano2017and}, we conduct a large-scale empirical but parallel study where we focus on a different scope, a much wider set of code TDIs, and different research questions. It is important to clarify that the study by Tufano et~al.~\cite{tufano2017and} uses the code smells by Fowler et~al.~\cite{Fowler1999} whereas we study the code smells as defined and detected by SonarQube, although, of course, there is some extent of overlap among the two code smells sets.

In addition to code smells, we investigate the survivability of bugs and vulnerabilities as defined by SonarQube. Bugs are of interest since they refer to coding violations that do not affect the internal quality of the system, as code-smells, but are coding violations that impact the external quality of the developed system~\cite{Kruchten2019}. Understanding how they are treated and how much time they survive in the system can help us understand how different types of TDI are mitigated and consider these findings to plan TD more efficient mitigation actions. Vulnerabilities are related to security issues and are usually discovered after some time has passed, making them different from code smells. Therefore, studying their survivability can give us more insight into how they are treated during the development. 

This sample study aims at assessing the survivability of code TDIs in large-scale industrial and open-source systems by using robust statistical tests. We conduct an empirical study to investigate further how long TDIs in codebases survive. To achieve this goal, we conduct a longitudinal study of two industrial systems and 31 open-source systems from the Apache Foundation. We  analysed 133,670 code TDIs in 193,196 commits in total to investigate \textit{the survivability of different code TDI types: bugs, code smells, and vulnerabilities.} 

More specifically, this study aims at answering the following research question:
\begin{itemize}
    \item \textbf{RQ.} \textit{What are the differences in the survivability of the types of code TDI, namely bugs, code smells, and vulnerabilities?}
\end{itemize}

This paper makes the following contributions: 

\begin{itemize}
    \item Provides insights about the survivability of code TDIs (bugs, code smells, and vulnerabilities);   
    \item Provides a replication package for reproducible results and analysis by other researchers\footnote{\url{https://github.com/ehsanzabardast/code_tdi_survivability}}.
\end{itemize}

The rest of this article is organised as follows: Section~\ref{sec:related_work} summarises the related works of studies. Section~\ref{sec:research_methododology} describes the data collection and analysis process. The results are described in Section~\ref{sec:results}. Section~\ref{sec:discussion} provides a discussion of the implications of the results. Lastly, the threats to validity  and conclusions of this study are presented in Section~\ref{sec:threats_to_validity} and Section~\ref{sec:conclusions} respectively.

%% file: sections/02_relatedwork.tex
\section{Related Work}\label{sec:related_work}

Several studies have focused on the detection and understanding of when and how the introduction of specific types of code TDI (i.e., code smells) impact software maintenance and quality (e.g.,~\cite{arcoverde2011understanding, peters2012evaluating}). These studies use historical data to evaluate the lifespan of code smells and help improve maintenance activities and code quality. Moreover, the empirical analysis of the evolution of code smells during the product life-cycle has been addressed in research studies~\cite{arcoverde2011understanding,lozano2007assessing}. Chatzigeorgiou and Manakos~\cite{chatzigeorgiou2010investigating} investigated the presence and evolution of code smells through an exploratory analysis of past versions of a software system. With a focus on when and how code smells are introduced and removed from software systems, the authors examined the evolution of three types of code smells in two open-source systems. Their results indicated that most code smells last as long as the software system operates, and refactoring does not necessarily eliminate code smells. This is not unusual as a study by Peters and Zaidman~\cite{peters2012evaluating} showed that developers, although being very aware of the presence of code smells in their code, tend to ignore the impact of those code smells on the maintainability of their code. This observation ignited interest in assessing the impact of code smells on maintenance activities within the research community, and several empirical studies have been conducted since then. Marcilio et~al.~\cite{marcilio2019static} investigated the usage of an automatic static analysis tool (ASAT), i.e., Sonarqube, to examine its usage by the developers. They report that practitioners can benefit from using ASATs if they are properly configured, i.e., using relevant rules. Their results show that only $8.76\%$ of the code TDIs are \textit{fixed} from the detected code TDIs among which code smells and major issues are more prevalent.

A study by Yamashita and Moonen~\cite{yamashita2012code} consists of an empirical study of four Java-based systems before entering the maintenance phase and observing and interviewing 14 developers who maintained the systems. They identified 13 maintainability factors that are impacted by code smells. Further empirical studies by the same authors (Yamashita and Moonen~\cite{yamashita2013exploring}), focused on the interaction between code smells and their effect on maintenance effort, revealed that some inter-smell relations were associated with problems during maintenance and some inter-smell relations manifested across coupled artefacts. A study by Sj{\o}berg et~al.~\cite{sjoberg2012quantifying} demonstrated that the effect of code smells on maintenance effort was limited. Digkas et~al.~\cite{digkas2017evolution} investigate the evolution of TD in open-source systems in the Apache ecosystem over time. Their results suggest that TD increases monotonically over time in most of the investigated systems.

More studies have recently empirically analysed different types of technical debt items in several other software projects intending to understand the evolution of code smells, when and how they are introduced into systems. Tufano et~al.~\cite{tufano2015and} conducted an extensive empirical study on 200 open-source systems and investigated when bad smells are introduced. Comprehensive analysis of over 0.5M commits and manual analysis of 9164 smell-introducing commits revealed that smells are not introduced during evolutionary tasks. A further study by the same authors~\cite{tufano2017and} contradicts common wisdom, showing that the majority of code smells are introduced when an artefact is created and not during the evolution process where several changes are made to software artefacts. They also observed that 80\% of smells are not removed, and they survive as long as the system functions, confirming previous results by Chatzigeorgiou and Manakos~\cite{chatzigeorgiou2010investigating}. 

Additionally, some research has been conducted on how much attention is dedicated to technical debt items such as code smells, bugs, and others by software developers and how these TDIs are resolved during the evolution process of a software system. A recent work by Digkas et~al.~\cite{digkas2018developers} analysed the life cycles of code TDIs in several open-source systems. The authors reported a case study focusing on the different types of code TDIs fixed by developers and the amount of technical debt repaid during the software evolution process. The study analysed the evolution (weekly snapshots) of 57 Java open-source software systems under the Apache ecosystem. The study revealed that allocating resources to fix a small subset of the issue types contributes towards repaying the technical debt. Similarly, the recent study of Saarim{\"a}ki et~al.~\cite{saarimaki2019diffuseness} aimed to comprehend the diffuseness of TD types, how much attention was paid to TDIs by developers and how severity levels of TDIs affected their resolution. The authors observed that code smells are the most introduced TDIs, and the most severe issues are resolved faster. To understand the needs of software engineers with regards to technical debt management, Arvanitou et~al.~\cite{arvanitou2019monitoring} surveyed 60 software engineers from 11 companies. The authors observed that developers were mostly concerned with understanding the underlying problems existing in source code, whereas managers cared most about financial concepts.

Prior studies have, in general, focused on studying the introduction and evolution of code smells in open-source software systems. To the best of our knowledge, only the work by Digkas et~al.~\cite{digkas2018developers} and our previous work (\cite{zabardast2020}) analysed TD repayment. Digkas et al. \cite{digkas2018developers} analysed  TD repayment by focusing on a broader set of TDIs, but with a coarse granularity (weekly snapshots) whilst we analyse the effect at commit level. In our previous work~\cite{zabardast2020}, we analysed the impact of different activities on TD, i.e., whether each activity contributed to the accumulation or repayment of TD, whilst in this paper, we focus on TDIs. We present a statistical model on the survivability of code TDIs.

This study aims to extend the findings of previous studies in literature, especially the study by Tufano et~al.~\cite{tufano2017and}. We aim to verify and complement the results obtained in the study of Tufano et~al.~\cite{tufano2017and} with regards to the survivability aspects by extending the scope of the analysis considering a bigger set of code smells and other categories of code TDI that can appear in code-bases such as \textit{bugs}, and \textit{vulnerabilities}. We thus analyse a similar set of code TDI as the one considered by Digkas et~al.~\cite{digkas2018developers} by focusing on individual commits.

%% file: sections/03_research_methodology.tex
\section{Research Methodology}\label{sec:research_methododology}

We have conducted a sample study to address the research questions defined in Section~\ref{sec:introduction}. The purpose of a sample study is to ``study the distribution of a particular characteristic in a population (of people or systems), or the correlation between two or more characteristics in a population.~\cite{stol2018abc}''  In the subsections below, we provide details on the data collection and analysis.

\subsection{Context Selection}\label{sec:context_selection}

We selected two industrial systems for this study and 31 open source systems (OSS) from the Apache Software Foundation, all developed in \textsc{Java}. We have collected the data for this study collaborating with two companies that work in the areas of communication technology (Industrial 1 system) and banking and financial services (Industrial 1 system). The industrial systems were selected by convenience and because they are long-lived, large-scale systems that are still in production and continuously evolving and have the following characteristics: 

\begin{itemize}
    \item \textbf{Industrial 1.} The selected system has over one million lines of code and has been under development for over ten years. We analysed 9,331 commits that contained 33,974 code TDIs. The developing company of this case is a large-size company (enterprise size).
    \item \textbf{Industrial 2.} The selected system has over 60,000 lines of code and has been under development for over four years. We analysed 8,414 commits that contained 1,398 code TDIs. The developers are using SonarQube in this system. The developing company of this case is a medium-size company (enterprise size).
\end{itemize}

The companies were selected by convenience and availability. The partner companies are mature in their development practices and have well-established, successful products. They are interested in continuously improving their products and development life-cycles, which turns into their willingness on participating in studies like this. All the collaborating companies work on developing software-intensive products and services.

Table~\ref{tab1:case_inf} summarises the historical information of the OSS analysed systems. All the OSS systems are hosted in \texttt{git} repositories\footnote{\url{https://github.com/apache}} and the SonarQube data was collected through a web API available online\footnote{\url{https://sonarcloud.io/organizations/apache/projects}}. The data collection for the industrial cases was performed using the corresponding SonarQube API in their on-site SonarQube installations.

\input{tables/tab1.case.inf}

\subsection{Using SonarQube for Code TDI Detection}\label{sec:issue_detection}

While there are alternative tools to detect the TDIs in the codebase of a system such as Codacy\footnote{\url{https://www.codacy.com/}} and PMD source code analyser\footnote{\url{https://pmd.github.io/pmd-6.17.0/index.html}}, we decided to use SonarQube\footnote{\url{https://www.sonarqube.org/}} because it is widely used in both industrial and open-source systems~\cite{saarimaki2019accuracy} and has been used in other research studies, e.g., Zabardast et~al.~\cite{zabardast2020}, Digkas et~al.~\cite{digkas2018developers}, and Guaman et~al.~\cite{guaman2017sonarqube}. SonarQube, similar to other static analysis tools, parses the code base, and builds a model for each commit being analysed. 

SonarQube classifies code TDIs into three types of \textit{Bugs}, \textit{Code Smells}, and \textit{Vulnerabilities}. SonarQube's definitions\footnote{\url{https://docs.sonarqube.org/latest/user-guide/issues/}} for these types are presented below with one example for each type\footnote{All the SonarQube rules can be browsed at \url{https://rules.sonarsource.com}}. 

\begin{itemize}
\item \textbf{Bug}: SonarQube defines bugs as a coding error that will break your code and needs to be fixed immediately. It is important to clarify that SonarQube \textit{bugs} are not what it is reported in issue tracking systems, but rather detected through its static analysis. SonarQube's definition contrasts with the most spread definition of \textbf{Bug}: a synonym of fault, which is a manifestation of an error in the software, such an incorrect step, process, or data definition in a computer program~\cite{ISO/IEC/IEEE2010}.  
\item \textbf{Code Smell}: SonarQube defines Code Smells as a maintainability issue that makes your code confusing and difficult to maintain. \textbf{Code Smells} are commonly defined as surface indications of deeper problems in the system~\cite{Fowler1999,fowler2018refactoring}, and they are ``sniffable'' at code level~\cite{Fowler2006}. Code-Smells are sometimes also referred to as Code Anti-Patterns, which are common re-occurring solutions to a problem, which generate negative consequences~\cite{Brown1998}.
\item \textbf{Vulnerability}: SonarQube defines Vulnerabilities as a point in your code that's open to attack. \textbf{Vulnerabilities} are usually referred as a weakness in an information system, system security procedures, internal controls, or implementation that could be exploited by a threat source~\cite{NIST2012}. A vulnerability ``does not cause harm in itself as there needs to be a threat present to exploit it~\cite{ISO/IEC2018}''.
\end{itemize} 

We want to emphasise that SonarQube's definitions of the code TDI types should be considered in its own context. In this paper, when we refer to the above-mentioned TDI types, we refer to SonarQube definitions.

The rest of this subsection is dedicated to providing examples for each type of code TDI from SonarQube.

\begin{displayquote}
\noindent\textbf{Bug - Regex lookahead assertions should not be contradictory}

Lookahead assertions are a regex feature that makes it possible to look ahead in the input without consuming it. It is often used at the end of regular expressions to make sure that substrings only match when they are followed by a specific pattern.

However, they can also be used in the middle (or at the beginning) of a regex. In that case there is the possibility that what comes after the lookahead does not match the pattern inside the lookahead. This makes the lookahead impossible to match and is a sign that there’s a mistake in the regular expression that should be fixed.

\textit{Noncompliant Code Example:}

\begin{lstlisting}
Pattern.compile("(?=a)b"); // Noncompliant, the same character can't be equal to 'a' and 'b' at the same time
\end{lstlisting}

\textit{Compliant Solution:}

\begin{lstlisting}
Pattern.compile("(?<=a)b");
Pattern.compile("a(?=b)");
\end{lstlisting}

\noindent\textbf{Code Smell - Methods returns should not be invariant}

When a method is designed to return an invariant value, it may be poor design, but it should not adversely affect the outcome of your program. However, when it happens on all paths through the logic, it is surely a bug.

This rule raises an issue when a method contains several return statements that all return the same value.

\textit{Noncompliant Code Example:}

\begin{lstlisting}
int foo(int a) {
  int b = 12;
  if (a == 1) {
    return b;
  }
  return b;  // Noncompliant
}
\end{lstlisting}

\noindent\textbf{Vulnerability - Server certificates should be verified during SSL/TLS connections}

Validation of X.509 certificates is essential to create secure SSL/TLS sessions not vulnerable to man-in-the-middle attacks.

The certificate chain validation includes these steps:

\begin{itemize}
    \item The certificate is issued by its parent Certificate Authority or the root CA trusted by the system.
    \item Each CA is allowed to issue certificates.
    \item Each certificate in the chain is not expired.
\end{itemize}

This rule raises an issue when an implementation of X509TrustManager is not controlling the validity of the certificate (ie: no exception is raised). Empty implementations of the \texttt{X509TrustManager} interface are often created to disable certificate validation. The correct solution is to provide an appropriate trust store.

\begin{lstlisting}
class TrustAllManager implements X509TrustManager {

    @Override
    public void checkClientTrusted(X509Certificate[] chain, String authType) throws CertificateException {  // Noncompliant, nothing means trust any client
    }

    @Override
    public void checkServerTrusted(X509Certificate[] chain, String authType) throws CertificateException { // Noncompliant, this method never throws exception, it means trust any server
        LOG.log(Level.SEVERE, ERROR_MESSAGE);
    }

    @Override
    public X509Certificate[] getAcceptedIssuers() {
        return null;
    }
}
\end{lstlisting}

\end{displayquote}

\subsection{Data Analysis}\label{sec:data_analysis}

\begin{figure*}
\centering
  \includegraphics[width=0.7\textwidth]{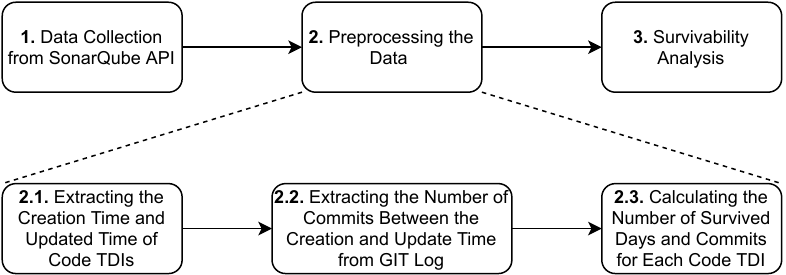}
\caption{Data analysis process.}
\label{fig:data_analysis}   
\end{figure*}

To understand the survivability of each code TDI's in the studied systems, we analyse the existence of code TDIs from their introduction in the codebase until they are marked as ``closed''. Similarly, the code TDIs that are still remaining in the system are marked as ``open''. In our analysis, we include \textit{all} the code TDIs present in the system, i.e., we include the code TDIs marked as closed and open. Figure~\ref{fig:data_analysis} summarises the data analysis procedure for this study. We use the collected data from SonarQube API (Step 1). We process the preprocess of the data to extract the number of survived days and commits for each code TDI (Step 2). The processed data is used for the survival analysis (Step 3). The details of the analysis are provided in the rest of this subsection.

We use the creation time and the updated time -- when the TDI was closed -- of each TDI in the collected data. The extracted information is used to calculate the number of survived days and the number of survived commits for each case.

We use \textit{The Number of Survived Days} and \textit{The Number of Survived Commits} similar to the study designed by Tufano et~al.~\cite{tufano2017and} as complementary metrics to capture the views of code TDI survivability. Considering these metrics individually might be misleading since a project can be inactive for months (i.e., nothing committed for a while). The processed data will contain code TDI with no removal time, i.e., the code TDIs that are not marked as ``closed'' in the collected data. To interpret this, we use the same approach as \cite{tufano2017and} and mark such code TDI as ``\textit{Censored Data}''.

We analyse the data using Survival analysis. Survival analysis is a statistical method that analyses and models the duration of events until other events happen~\cite{miller2011survival}, i.e., code TDI removal in this case. The survival function for code TDI, $S(t)=Pr(TDI>t)$ indicates that a code TDI exists longer than a specific time $t$. The survival analysis creates a Survivability Model based on historical data. Using this model, we can generate survival curves illustrating the survival probability as a function of time. The model can handle both \textit{complete data} (observations with an ending event) and \textit{incomplete data} (observations without an ending event) if the data is marked properly. We create survival models based on the number of survived days and the number of survived commits for each type of code TDI, namely bugs, code smells, and vulnerabilities. The analysis is done in \texttt{R} using the \texttt{survival}\footnote{\url{https://cran.r-project.org/web/packages/survival/index.html}} and \texttt{survminer}\footnote{\url{https://cran.r-project.org/web/packages/survminer/index.html}} packages. The \texttt{Surv} function is used to generate the survival model, and the \texttt{survfit} function is used to estimate survival curves. Additionally, similar to \cite{tufano2017and}, we utilise the Kaplan-Meier estimator~\cite{kaplan1958nonparametric} in the analysis to estimate the removal time for the \textit{incomplete} observations, therefore we also included ``censored'' data-points.

%% file: tables/tab1.case.inf.tex
\begin{center}
\begin{table}[t]%
\centering
\caption{General description of the analysed open-source systems.\label{tab1:case_inf}}%
\begin{tabular*}{500pt}{@{\extracolsep\fill}ccccccc@{\extracolsep\fill}}
\toprule
&&& \multicolumn{4}{@{}c@{}}{\textbf{Number of}} \\\cmidrule{4-7}
&\textbf{System Name}&\textbf{System Size}&\textbf{Commits}&\textbf{TDIs}&\textbf{Classes}&\textbf{Contributors} \\
\midrule
1& Ant (v 1.10.0)&12503&14698&9351&1321&53\\
2& Commons Compress (v 1.21)&26974&3133&772&377&56\\
3& Commons Geometry (v 1.1)&16483&431&99&355&10\\
4& CXF (v 3.5.0)&427463&16155&10000\tnote{$\dagger$}&7555&154\\
5& Groov (v 3.0.9)&189114&18181&10000\tnote{$\dagger$}&1724&307\\
6& Hadoop Ozone (v 1.2.0)&132255&3202&3531&2182&104\\
7& IoTDB Project (v 0.13.0)&120051&4611&2082&1685&94\\
8& Isis (Aggregator) (v 2.0.0)&165713&15676&5532&5232&40\\
9& Jackrabbit FileVault (v 3.5.1)&47692&8859&2053&3128&23\\
10& JSPWiki (v 2.11.0)&56647&8853&3598&618&13\\
11& Karaf (v 4.3.1)&123212&8482&6567&1582&141\\
12& PDFBox (v 3.0.0)&141971&9694&1876&1348&6\\
13& POI (version unspecified)&259765&10679&10000\tnote{$\dagger$}&3512&15\\
14& Ratis (v 2.2.0)&38622&1131&957&607&41\\
15& ServiceComb Pack (v 0.7.0)&17494&1568&614&458&55\\
16& ServiceComb Toolkit (v 0.3.0)&15262&237&331&639&5\\
17& Shiro (v 2.0.0)&32723&2125&1964&724&46\\
18& Sling - CMS (v 1.0.5)&11409&894&52&287&9\\
19& Sling Distribution Core (v 0.4.3)&14185&483&621&267&9\\
20& Sling Launchpad Integration Tests (v 11)&16180&483&1901&167&17\\
21& Sling Resource Resolver (v 1.7.11)&10377&710&444&101&24\\
22& Sling Scripting JSP (v 2.5.5)&27233&275&1929&124&12\\
23& Incubator Tamaya (Retired) (v 0.4)&19056&1618&702&460&10\\
24& Dolphin Scheduler (v 1.3.6)&58979&4441&2135&912&175\\
25& Gateway (v 1.6.0)&76495&2455&2241&1479&54\\
26& Hop Orchestration Platform (v 1.0)&483430&1572&10000\tnote{$\dagger$}&3240&20\\
27& Jmeter (v 5.5)&117352&17331&4362&1386&35\\
28& Openmeetings (v 7.0.0)&99052&3121&1439&584&12\\
29& PLC4X (v 0.9.0)&57162&3740&2491&1007&43\\
30& Roller (v 6.1.0)&66248&4549&3318&610&12\\
31& Struts 2 (v 2.6)&110941&6064&8142&2060&51\\
\midrule
&\textbf{Totals}&2992043&175451&109104&45731&-\\
\bottomrule
\end{tabular*}
\begin{tablenotes}
\item[$\dagger$] The limit for fetching the data from API is 10000 code TDIs.
\end{tablenotes}
\end{table}
\end{center}

%% file: sections/04_results.tex
\section{Results}\label{sec:results}

In this section, we report the results of our study. We focus on the three types of code TDI (i.e., code smells, bugs, and vulnerabilities), as done in \cite{digkas2018developers}. The raw experimental results, data sets extracted from open-source systems and the source code for implementing the experiments is provided in our replication kit online\footnote{\url{https://github.com/ehsanzabardast/code_tdi_survivability}}.

\begin{figure*}
\centering
  \includegraphics[width=0.6\textwidth]{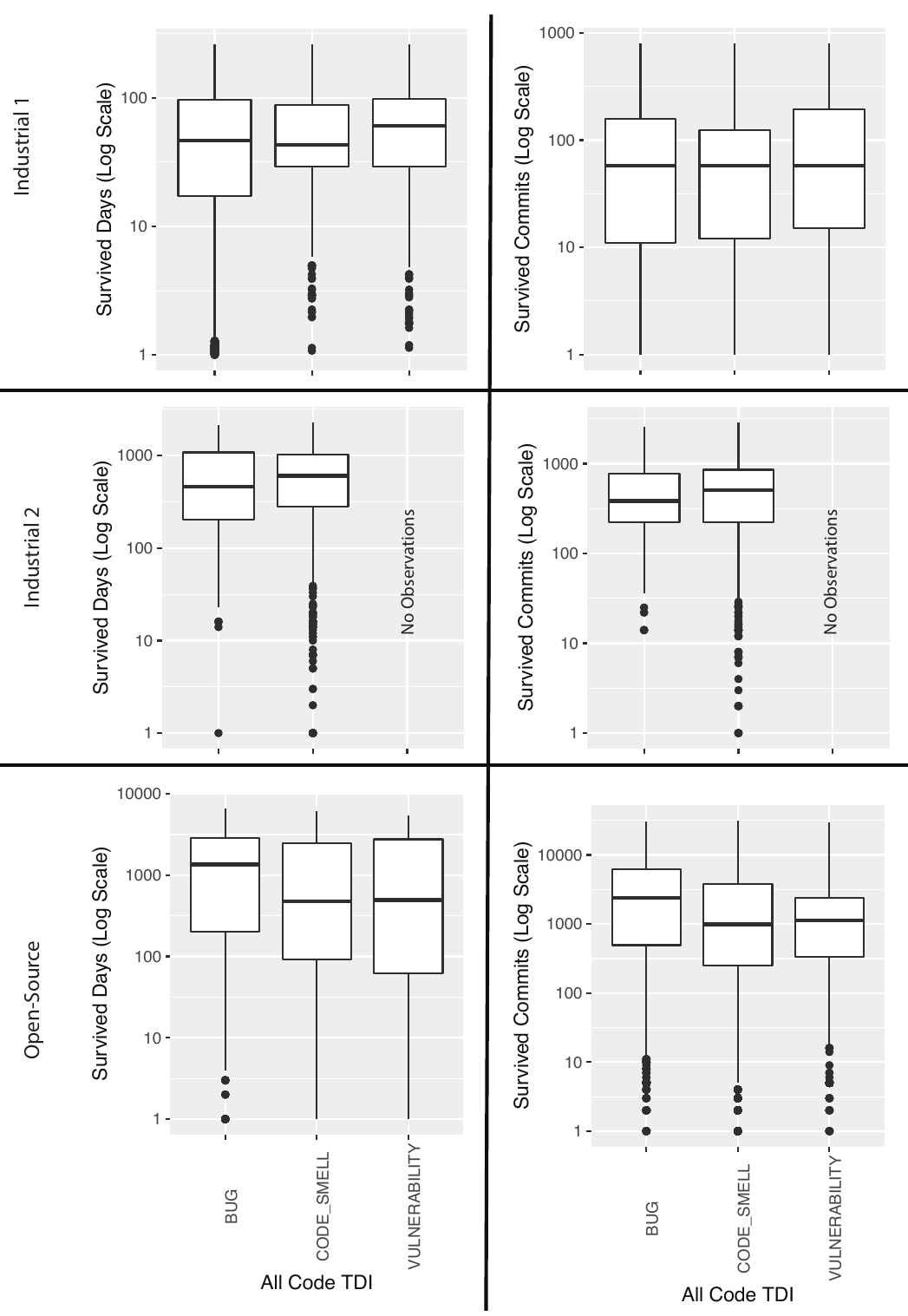}
\caption{Distribution of the number of days (left) and commits (right) that code TDIs survived. The first row belongs to the industrial system and the second row belongs to the open-source systems. Note that the data presented in this figure is representative only for the cases with terminal event occurred.}
\label{fig:02_all_issues_box}   
\end{figure*}

To address the research question, we analysed the survivability of the code TDIs collected from two industrial systems and 31 open-source systems from the Apache Foundation. We have analysed the survivability both in terms of the survived days and survived commits. The breakdown of identified code TDIs per system per type is summarised in Table~\ref{tab1:tdi_count}.
Figure~\ref{fig:02_all_issues_box} illustrates the box plots for the distribution of the survived days (left) and commits (right) for the detected code TDIs. The first and second rows belong to the industrial systems, and the third row belongs to the open-source systems. The plots are presented on a log scale to make the representation easier to read. We observe that the medians of the distributions are not significantly different across different types of code TDI for the number of survived days and the number of survived commits. However, they are much lower in the industrial systems for all types of code TDIs.

\input{tables/tab.tdi.count}

In the case of Industrial 2, SonarQube did not detect any vulnerabilities (marked as ``No Observations'' in Figure~\ref{fig:02_all_issues_box}). For this particular system, we have analysed the last two years only. The fact that vulnerabilities are often reported only until some time has passed might explain why we have not found any vulnerabilities in this particular case. Similarly, in Table~\ref{tab:issues_descriptives}, the presented descriptive statistics, for the case of industrial 2, are only for code smells and bugs, and they should be interpreted keeping this in mind. 

Table~\ref{tab:issues_descriptives} summarises the descriptive statistics for the number of survived days and commits for all systems. The table distinguishes between the industrial systems and open-source systems for each row labelled as \textit{Industry} and \textit{Open-Source}. We use the median as the measure of central tendency to minimise the effect of outliers~\cite{leys2013detecting}. We use median values as the references of analysis, as summarised in Table~\ref{tab:issues_descriptives}. Note that the data presented in Figure~\ref{fig:02_all_issues_box} and Table~\ref{tab:issues_descriptives} are only representative of the cases with terminal event occurred.

We observe that the majority of the code TDIs detected in the industrial 1 are removed before the $43^{rd}$ day (before $58^{th}$ commit), and the majority of the code TDIs detected in the industrial 2 are removed before the $580^{th}$ day (before $496^{th}$ commit). In open-source systems, the majority of the code TDIs are removed before the $398^{th}$ day (before $722^{nd}$ commit). 

\input{tables/tab2.tdi.descriptives}

\begin{figure*}
\centering
  \includegraphics[width=0.8\textwidth]{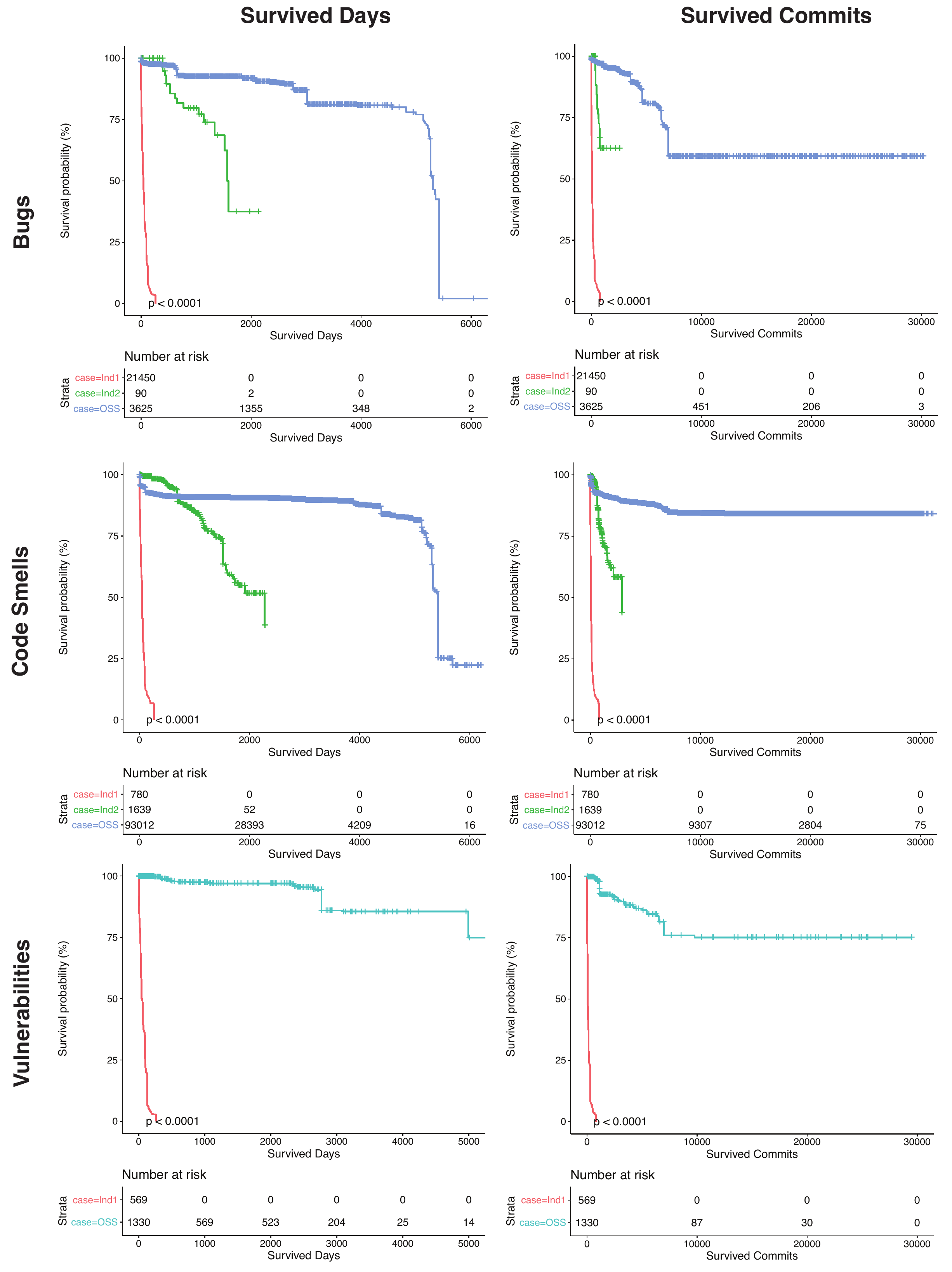}
\caption{Distribution of the number of days (left) and commits (right) that code TDIs survived in all systems.}
\label{fig:03_all_issues_surv}   
\end{figure*}

Figure~\ref{fig:03_all_issues_surv} illustrates the calculated survival probability curves for all the systems (survived days on the left and survived commits on the right). The first row belongs to the industrial 1 system, the second row belongs to the industrial 2 system, and the third row belongs to the open-source systems.

Our first observation is that the survivability of code TDIs vary in the systems under investigation, both in terms of the number of days and the number of commits. Considering these calculated probability curves, the survival probability of code TDIs in terms of survived days is highest in open-source systems. In contrast, the survival probability of code TDIs in industrial systems is lower. Having a higher survivability probability in terms of days means that for the same number of days, code TDIs in open-source systems have a higher probability of surviving as compared to the probability of a code TDI surviving with the same number of days in the industrial systems. In other words, we have observed that in the analysed systems, code TDIs tend to be removed faster, in terms of the number of days, in industrial settings.

We also observe that the survival probability of code TDIs in terms of survived commits is higher in open-source systems. Having a higher survivability probability in terms of commits means that for the same number of commits, code TDIs in open-source systems have a higher probability of surviving as compared to the probability of a code TDI surviving with the same number of commits in the industrial systems. In other words, we have observed that in the analysed systems, code TDIs tend to be removed faster, in terms of the number of commits, in industrial settings.

We use $100$ days as the point of reference to present the results of survivability models. However, instead of using $10$ commits as in Tufano et al. ~\cite{tufano2017and} work, we use $100$ commits in our analysis as the point of reference. This is owing to the fact that the $10$ commit threshold might be too short for our data sets, and the results might turn inconclusive since the probabilities of survivability for ten days are very high and very similar among systems.

We observe that, in general, code TDIs are removed faster in the investigated industrial system as compared to the investigated open-source systems. Table~\ref{tab3:surv.numbers} summarises the probability of code TDIs surviving up to $100$ days and commits for \textit{Bug}, \textit{Code Smell}, and \textit{Vulnerability} separately for the industrial system and open-source systems. Note that there was no observation as \textit{vulnerability} in the Industrial 2. Therefore we cannot be presenting any survival probability in Table~\ref{tab3:surv.numbers} for Industrial 2.


\input{tables/tab3.surv.numbers}

From the survivability models for all code TDIs in the investigated systems in Table~\ref{tab3:surv.numbers}, we observe that:
\begin{itemize}
    \item \textbf{In the industrial 1 system:} There is a higher chance for code TDIs to be removed before $100$ days ($100$ commits), i.e., many of the code TDIs are removed relatively soon.
    \begin{itemize}
        \item \textit{Bug:} There is a $18.476\%$ chance that the investigated code TDIs survive until $100$ days ($38.312\%$ chance that the investigated code TDIs survive until $100$ commits).
        \item \textit{Code Smell:} There is a $13.850\%$ chance that the investigated code TDIs survive until $100$ days ($37.180\%$ chance that the investigated code TDIs survive until $100$ commits).
        \item \textit{Vulnerability:} There is a $23.200\%$ chance that the investigated code TDIs survive until $100$ days ($39.890\%$ chance that the investigated code TDIs survive until $100$ commits).
    \end{itemize}
    \item \textbf{In the industrial 2 system:} There is a lower chance for code TDIs to be removed before $100$ days ($100$ commits).
    \begin{itemize}
        \item \textit{Bug:} There is a $100\%$ chance that the investigated code TDIs survive until $100$ days ($98.360\%$ chance that the investigated code TDIs survive until $100$ commits).
        \item \textit{Code Smell:} There is a $99.410\%$ chance that the investigated code TDIs survive until $100$ days ($99.320\%$ chance that the investigated code TDIs survive until $100$ commits).
        \item \textit{Vulnerability:} There are not enough observations to calculate survival probabilities.
    \end{itemize}
    \item \textbf{In open-source systems:} Similar to the industrial systems, there is a higher chance for code TDIs to be removed before $100$ days ($100$ commits), i.e., many of the code TDIs are removed relatively soon but if they are not removed, they can stay in the system for as long as the system is in production.
    \begin{itemize}
        \item \textit{Bug:} There is a $91.838\%$ chance that the investigated code TDIs survive until $100$ days ($92.619\%$ chance that the investigated code TDIs survive until $100$ commits).
        \item \textit{Code Smell:} There is a $94.470\%$ chance that the investigated code TDIs survive until $100$ days ($95.227\%$ chance that the investigated code TDIs survive until $100$ commits).
        \item \textit{Vulnerability:} There is a $99.381\%$ chance that the investigated code TDIs survive until $100$ days ($99.683\%$ chance that the investigated code TDIs survive until $100$ commits).
    \end{itemize}
\end{itemize}

We analysed the different types of code TDIs individually for the industrial system and open-source systems. Individual box plots for the distribution of the survival days (left) and commits (right) for the different types of code TDI detected in the industrial systems, and open-source systems are presented in Figure~\ref{fig:02_all_issues_box}. We observe that the median of the distribution for bug, code smell, and vulnerability is significantly higher in the open-source systems when compared to the industrial systems both for the number of survived days and the number of survived commits.

%% file: tables/tab.tdi.count.tex
\begin{table*}
\caption{The breakdown of identified code TDIs per system per type.}
\begin{center}
\begin{tabular}{ccccc}
\toprule
\textbf{TDI Type} & \textbf{Industrial 1} & \textbf{Industrial 2} & \textbf{Open-Source} & \textbf{Total} \\
\midrule
\textit{Bugs} & 31,808 & 90 & 3,625 & 35,523 \\
\textit{Code Smells} & 1,266 & 1,639 & 93,012 & 95,917 \\
\textit{Vulnerabilities} & 900 & 0 & 1330 & 2,230 \\
\textbf{Total} & 33,974 & 1,729 & 97,967 & 133,670\\
\bottomrule
\end{tabular}
\label{tab1:tdi_count}
\end{center}
\end{table*}

%% file: tables/tab2.tdi.descriptives.tex
\begin{center}
\begin{table}[t]%
\centering
\caption{Descriptive statistics for the number of survived days and commits that code TDIs survived in the systems. Note that the data presented in this figure is only representative of the cases with terminal event occurred.}
\begin{tabular}{ccccccc}
\toprule
\textbf{Case} & \textbf{Min} & \textbf{1stQu.} & \textbf{Median} & \textbf{Mean} & \textbf{3rdQu.} & \textbf{Max} \\
 \cline{2-7} \\
 & \multicolumn{6}{c}{\textbf{Survived Days}} \\

\textit{Industry 1} - All Code TDI & 0 & 12 & 43 & 58.70 & 95 & 262 \\
\textit{Industry 2} - All Code TDI & 0 & 221 & 580 & 678.01 & 1,007 & 2,283 \\
\textit{Open-Source} - All Code TDI & 0 & 21 & 398 & 1,255.80 & 2,370 & 6,656 \\
\midrule
\textit{Industry 1} - Bug & 0 & 11 & 43 & 58.37 & 96 & 262 \\
\textit{Industry 2} - Bug & 1 & 203 & 462 & 665.81 & 1,083 & 2,134 \\
\textit{Open-Source} - Bug & 0 & 168 & 1,287 & 1,662.78 & 2,767 & 6,656 \\
\midrule
\textit{Industry 1} - Code Smell & 0 & 15 & 43 & 62.32 & 83 & 252 \\
\textit{Industry 2} - Code Smell & 0 & 221 & 584 & 679.16 & 989 & 2,283 \\
\textit{Open-Source} - Code Smell & 0 & 18 & 398 & 1,238.73 & 2,338 & 6,211 \\
\midrule
\textit{Industry 1} - Vulnerability & 0 & 23 & 49 & 1,340.52 & 98 & 261 \\
\textit{Industry 2} - Vulnerability & - & - & - & - & - & - \\
\textit{Open-Source} - Vulnerability & 0 & 26 & 415 & 1,340.52 & 2,767 & 5,470 \\
\midrule \\
& \multicolumn{6}{c}{\textbf{Survived Commits}} \\
\textit{Industry 1} - All Code TDI & 1 & 11 & 58 & 131.72 & 155 & 795 \\
\textit{Industry 2} - All Code TDI & 1 & 200 & 496 & 603.29 & 813 & 2,881 \\
\textit{Open-Source} - All Code TDI & 1 & 68 & 722 & 3,154.40 & 3,492 & 31,690 \\
\midrule
\textit{Industry 1} - Bug & 1 & 11 & 58 & 131.11 & 158 & 795 \\
\textit{Industry 2} - Bug & 1 & 224 & 375 & 545.30 & 773 & 2577 \\
\textit{Open-Source} - Bug & 1 & 255 & 1,719 & 4,443.13 & 4,950 & 30,185 \\
\midrule
\textit{Industry 1} - Code Smell & 1 & 12 & 58 & 143.38 & 124 & 695 \\
\textit{Industry 2} - Code Smell & 1 & 198 & 498 & 606.80 & 813 & 2,881 \\
\textit{Open-Source} - Code Smell & 1 & 67 & 707 & 3,114.26 & 3,311 & 31,690 \\
\midrule
\textit{Industry 1} - Vulnerability & 1 & 15 & 58 & 138.74 & 193 & 790 \\
\textit{Industry 2} - Vulnerability & - & - & - & - & - & - \\
\textit{Open-Source} - Vulnerability & 1 & 186 & 1,134 & 2,448.88 & 2,376 & 29,493 \\
\bottomrule
\end{tabular}
\label{tab:issues_descriptives}
\end{table}
\end{center}

%% file: tables/tab3.surv.numbers.tex
\begin{table*}
\caption{The probability of code TDIs surviving up to $100$ days and commits.}
\begin{center}
\begin{tabular}{ccccc}
\toprule
\textbf{Case} & \textbf{Survival Probability} & \textbf{Standard Error} & \textbf{Lower $95\%$ CI} & \textbf{Upper $95\%$ CI} \\
 \cline{2-5} \\
 & \multicolumn{4}{c}{\textbf{Survived Days}} \\
\textit{Industry 1} - Bug & 18.476\% & 0.00265 & 0.17963 & 0.19002 \\
\textit{Industry 2} - Bug & 100.000\% & 0 & 1 & 1 \\
\textit{Open-Source} - Bug & 91.838\% & 0.00474 & 0.90913 & 0.92773 \\
\midrule
\textit{Industry 1} - Code Smell & 13.850\% & 0.01240 & 0.11620 & 0.16500 \\
\textit{Industry 2} - Code Smell & 99.410\% & 0.00221 & 0.98982 & 0.998448 \\
\textit{Open-Source} - Code Smell & 94.470\% & 0.00084 & 0.94306 & 0.94635 \\
\midrule
\textit{Industry 1} - Vulnerability & 23.200\% & 0.01770 & 0.19980 & 0.26940  \\
\textit{Industry 2} - Vulnerability & - & - & - & -  \\
\textit{Open-Source} - Vulnerability & 99.381\% & 0.00236 & 0.98920 & 0.99844  \\
\midrule \\
& \multicolumn{4}{c}{\textbf{Survived Commits}} \\
\textit{Industry 1} - Bug & 38.312\% & 0.00332 & 0.37667 & 0.38968 \\
\textit{Industry 1} - Bug & 98.360\% & 0.01630 & 0.95230 & 1.0000 \\
\textit{Open-Source} - Bug & 92.619\% & 0.00447 & 0.91747 & 0.93499 \\
\midrule
\textit{Industry 1} - Code Smell & 37.180\% & 0.01730 & 0.33940 & 0.40730 \\
\textit{Industry 2} - Code Smell & 99.320\% & 0.00238 & 0.98862 & 0.99793  \\
\textit{Open-Source} - Code Smell & 95.227\% & 0.00078 & 0.95074 & 0.95380 \\
\midrule
\textit{Industry 1} - Vulnerability & 39.890\% & 0.02050 & 0.36070 & 0.44130  \\
\textit{Industry 2} - Vulnerability & - & - & - & -  \\
\textit{Open-Source} - Vulnerability & 99.683\% & 0.00158 & 0.99374 & 0.99994  \\
\bottomrule
\end{tabular}
\label{tab3:surv.numbers}
\end{center}
\end{table*}

%% file: sections/05_discussion.tex
\section{Discussion}\label{sec:discussion}

\subsection{General Findings}
Our empirical analyses reveal the unpredictable nature of the survivability of code TDIs in software systems. The survivability of the issues ranges from $0$ to $2,283$ days for the industrial systems and  $0$ to $6,656$ days for open-source systems.

The median for the industrial system stay below $100$ days; therefore, the TD mitigation strategies to remove code TDIs seem to be more effective in industrial systems as compared to open-source systems. Our results also suggest that, for open-source systems, when code TDIs remain in the system after $100$ days, they can survive for a long time in the system. Similar behaviour can be observed when analysing the number of survived commits, which ranges from $1$ to $2,881$ commits for the industrial systems and $1$ to $31,690$ commits for open-source systems. Our results are aligned with the findings of Digkas~et~al. \cite{digkas2017evolution}. In their paper, the authors present the monotonic upward trend growing TD over time in open source systems from the Apache Software Foundation. The monotonic upward trend is a sign that the code TDIs survive for longer periods in the systems under investigation.

TD mitigation strategies seem to be more effective in removing TDIs in industrial settings in terms of the number of survived commits as well. As discussed before, regardless of the system under study, code TDIs that survive past the ${100}^{th}$ day threshold might survive much longer, confirming the finding in the study by Tufano et~al. \cite{tufano2017and}. 

However, we observed patterns that contrast what was found in previous research studies. We found that:

\begin{itemize}
    \item After $100$ days:
    \begin{itemize}
        \item The survival probability for the industrial 1 system is as follows: \textit{Bug} $17.48\%$ - \textit{Code Smell} $13.85\%$ - \textit{Vulnerability} $23.20\%$.
        \item The survival probability for the industrial 2 system is as follows: \textit{Bug} $100.00\%$ - \textit{Code Smell} $99.41\%$.
        \item The survival probability for the open-source systems is as follows: \textit{Bug} $91.84\%$ - \textit{Code Smell} $94.47\%$ - \textit{Vulnerability} $99.38\%$.
    \end{itemize}
    \item After $100$ commits:
    \begin{itemize}
        \item The survival probability for the industrial 1 system is as follows: \textit{Bug} $38.31\%$ - \textit{Code Smell} $37.18\%$ - \textit{Vulnerability} $39.89\%$.
        \item The survival probability for the industrial 2 system is as follows: \textit{Bug} $98.36\%$ - \textit{Code Smell} $99.32\%$.
        \item The survival probability for the open-source systems is as follows: \textit{Bug} $92.62\%$ - \textit{Code Smell} $95.23\%$ - \textit{Vulnerability} $99.68\%$.
    \end{itemize}
\end{itemize}

The study by Tufano et~al.~\cite{tufano2017and} found that it was after $1000$ days when the survival probability achieved similar values. This might be owing to the fact that we have extended the scope of our analysis to include a much wider set of code smells (the study by Tufano et~al.~\cite{tufano2017and} only studied five code-smells), and we also analysed bugs and vulnerabilities, which were not in the scope of previous research studies. We hypothesise that some of the five code smells analysed in the previous study by Tufano et~al.~\cite{tufano2017and} might not be the priority for developers in their maintenance activities, in line with what is found in~\cite{Palomba2018}. Furthermore, we have observed a completely different behaviour when it comes to code-smells. Chatzigeorgiou and Manakos~\cite{chatzigeorgiou2010investigating} observed that code smells are never removed and stay as long as the software system operates. Similarly, Marcilio et~al.~\cite{marcilio2019static} observed that a low percentage ($8.76\%$) of the code TDIs, including bugs, code smells, and vulnerabilities, are removed, suggesting that not all code TDIs detected by SonarQube are relevant to the developers.

In the case of the industrial 1 system, we are aware that the development team put an emphasis on clean code practices. This might explain the faster rate of removal of code TDIs as illustrated in Figure~\ref{fig:03_all_issues_surv}. The results of our work were shared and discussed with the developer of the industrial 2 system. Throughout our discussions with the developers of the industry 2 system, they have informed us that their approach is to using SonarQube during the development. They do not use SonarQube to fix the existing problems, but they prevent new Code TDIs from arriving at the system.

\subsection{Implications of the Results}\label{sec:implications}

Technical Debt (TD) management has recently been the focus of attention in academic and industrial communities~\cite{rios2018tertiary}. The research on TD is in its initial phase, with researchers focusing on a few types of debt~\cite{li2015systematic,alves2016identification}. Empirical evaluation of TD management activities, especially the evidence from the software industry, is essential to shedding light on technical debt prioritisation activities~\cite{rios2018tertiary}. The results and the analysis methods provided by our study can help the research and industrial communities to better understand what the lifespan of different types of code TDIs is, not only code smells,  and invite researchers to further investigate TD prioritisation and the activities related to it. Moreover, analysing the survivability of code TDIs in a system can be performed on fine-grained code TDIs and not just the main types, e.g., specific code-smells in isolation. The survival analysis can help the developers in two ways. First, by helping them become aware of the survivability of existing code TDIs to repay them eventually. Second, by helping them prioritise action plans to prevent the accumulation of similar code TDIs similar to the case of Industrial 2 system's developers.

\textbf{\textit{SonarQube use and its impact on the survivability of the code TDI}}

The use of automatic static analysis tools has become popular in the last few years~\cite{marcilio2019static}. Given the fact that developers working on the Industrial 2 system have been using SonarQube during the last year of the development, we have examined if the use of such tools impacts the survivability of code TDIs by manually investigating the code TDIs that were created in the last year. It seems that the survivability of the code TDIs, both in terms of the number of survived days and commits, are not to be impacted by using SonarQube and follow a similar pattern to what is portraited in Figure~\ref{fig:03_all_issues_surv} for the Industrial 2 system. The code TDIs that were introduced in the last year during the period where developers used SonarQube have a similar number of survival days and commits.

\textbf{\textit{System type impacts the survivability of the code TDIs.} }

Our study suggests that the survival probability of the code TDIs, both in terms of the number of survived days and commits, varies throughout the software systems under investigation. By comparing the density distribution for all code TDIs of survived days with the density distribution for all code TDIs of survived commits presented in Figure~\ref{fig:02_all_issues_box}, we observe that the industrial and open-source systems have different distributions, but these distributions follow a similar trend. The code TDIs have a similar distribution in terms of the number of days but different survival probabilities in terms of the number of commits. This might be owing to the fact that the open-source systems have more frequent commits as compared to the industrial systems. Therefore, the code TDIs are addressed in later commits, whereas in industrial systems, the code TDIs tend to be addressed and resolved closer to the point when they are introduced in the system. 

\textbf{\textit{Code TDIs in the industrial systems are removed faster in terms of the number of commits.} }

Our analysis reveals that code TDIs survive longer in open-source systems as compared to industrial projects. A viable reason can be that the quality standards in the industrial systems prevent code TDIs from staying in the system for longer periods. 
There are other factors that might affect how long the code TDIs survive in different systems, such as the development practices put in place, the domain, the business model, the product maturity, and the expertise of the development team, and as discussed above, the usage of static analysis tools like SonarQube. The circumstances of each system might affect the survivability of the code TDIs, e.g., the industrial systems might have more rigorous development processes, including more stringent code reviews and test processes before code is pushed into production. 

Before drawing firm conclusions, each system should be analysed in isolation, considering additional factors. These factors might include the development process, developers' experience (in general and in the system), team's culture, product maturity, specific refactoring policies, developers' perception of whether the code contains code TDIs or not, and the willingness of developers to fix code TDIs.

%% file: sections/06_theats.tex
\section{Threats To Validity}\label{sec:threats_to_validity}
In this section, we present the potential threats to validity that might affect the results and findings of this study. We discuss below the threats to the construct, internal, and external validity of the study.

The main threat to validity is the \textit{Construct Validity}, i.e., the relationship between the theory and observation. The construct validity threats comprise of the errors and imprecision in measurement procedure adopted during the data collection process and whether the measurements actually reflect the construct being studied. We use SonarQube, a widely used tool for measuring TD, to detect code TDIs in the system. We also use the categories defined by SonarQube and employed in other research studies (e.g., \cite{digkas2018developers}), to categorise the types of code TDIs. We identify code TDIs in the systems using the default profile for \texttt{Java} by SonarQube. We acknowledge the problems that might arise due to the use of a particular tool, i.e., SonarQube, which includes the thresholds, measurements, and rules used to detect code TDIs and the possibility of having false positives and false negatives in the collected data.

Another threat to validity is regarding the detection of bugs and vulnerabilities. Bugs and vulnerabilities that are detected now might not have been detected when the systems were being developed, and there was no way for the developers to be aware of their existence. Therefore, no tool could warn the developers to remove the flaws when they were introduced.

\textit{Internal Validity} refers to the degree to which the presented evidence can support a cause and effect relationship within the context of the study.

Our results are based on the \texttt{R} packages used to calculate the survivability curves. Different implementations of the same survivability analysis might lead to obtaining different results. The fact that one of the systems (i.e., Industrial 1) was developed following the principles of clean-code~\cite{Martin2008} might have an influence on the results of the analysis of how code TDIs are removed, especially when it comes to the analysis of code smells and bugs being removed due to the development of new features.

Another threat to validity is regarding the detection of closed code TDIs. There is no way for us to detect the code TDIs that are marked as ``closed'' purposefully. There might be cases where code TDIs are marked as closed because a method or class is removed from the source code.

\textit{External Validity} refers to the degree to which the results can be generalised. In this study, we analysed five systems, which are mainly Java-based software systems, two industrial systems, and thirty-one open-source systems from the Apache Software Foundation. We limit the results of this study to the systems under investigation. We understand and acknowledge that the generalisability of the results is limited and the results are representative of the \textit{universe}, \textit{dimensions}, and \textit{configuration} covered in our study~\cite{nagappan2013diversity}. We can only claim that the results are applicable to the analysed commits in the systems under investigation.

We acknowledge that the removal of the code TDIs and the categories are highly dependent on each system, i.e., the development process, the developers' experience both in the software system project and their overall experience, the team's culture, the stage of development, and other factors that affect the survivability of code TDIs given a system \cite{tufano2017and}. Our analysis and findings are thus impacted by the system type under consideration.

%% file: sections/07_conclusions.tex
\section{Conclusions}\label{sec:conclusions}

In this paper, we present the results of a sample study on the survivability of code TDIs in software systems.
This paper presents a study on the change history of five software systems, and it aims to understand the survivability of code technical debt items (TDIs) in the codebase. Furthermore, this study aims to extend the results of prior studies by including smells and other types of code TDIs, such as bugs and vulnerabilities in the analysis code. Therefore, we have focused on examining and assessing the differences in survivability among the three categories of detected code TDI.

We have conducted a comprehensive empirical sample study using data from other software systems, including two large industrial software systems and 31 open-source systems from the Apache Foundation. As illustrated by the survival curves, most of the code TDI in the investigated systems are removed rather quickly, i.e., there is a $23.20\%$ chance that they survive until $100$ days. Any code TDI that survives this threshold has a higher probability of surviving longer in the system.

When the system type and commit activities are taken into account, the code TDIs in the systems which have a bigger size tend to have a longer survival duration. On the other hand, the code TDIs that survive past the median threshold tend to stay in the system for a long time.

Our findings open the door for further studies. Our results can be strengthened by digging into the other factors that affect the systems. Additionally, we believe replications are needed to strengthen the results and study whether the results might be generalizable to similar systems.

%% file: main.bbl
\begin{thebibliography}{10}
\providecommand \doibase [0]{http://dx.doi.org/}%

\bibitem{Lehman1979}
Lehman M. {On understanding laws, evolution, and conservation in the
  large-program life cycle}. {\it Journal of Systems and Software} 1979\string;
  1\string: 213--221.

\bibitem{Lehman1996}
Lehman M. {Laws of software evolution revisited}. {\it European Workshop on
  Software Process Technology} 1996\string: 108--124.

\bibitem{Kruchten2019}
Kruchten P, Nord R, Ozkaya I. {\it {Managing Technical Debt: Reducing Friction
  in Software Development}}.
\newblock Pearson .
\newblock 2019.

\bibitem{cunningham1993wycash}
Cunningham W. The WyCash portfolio management system. {\it ACM SIGPLAN OOPS
  Messenger} 1993\string; 4(2)\string: 29--30.

\bibitem{brown2010managing}
Brown N, Cai Y, Guo Y, et al. Managing technical debt in software-reliant
  systems. {\it Proceedings of the FSE/SDP workshop on Future of software
  engineering research} 2010\string: 47--52.

\bibitem{Kruchten2012}
Kruchten P, Nord RL, Ozkaya I. {Technical debt: From metaphor to theory and
  practice}. {\it IEEE Software} 2012\string; 29(6)\string: 18--21.

\bibitem{lim2012balancing}
Lim E, Taksande N, Seaman C. A balancing act: What software practitioners have
  to say about technical debt. {\it IEEE software} 2012\string; 29(6)\string:
  22--27.

\bibitem{Li2015}
Li Z, Avgeriou P, Liang P. {A systematic mapping study on technical debt and
  its management}. {\it Journal of Systems and Software} 2015\string;
  101\string: 193--220.

\bibitem{saarimaki2019diffuseness}
Saarim{\"a}ki N, Lenarduzzi V, Taibi D. On the diffuseness of code technical
  debt in Java projects of the apache ecosystem. {\it Proceedings of the Second
  International Conference on Technical Debt} 2019\string: 98--107.

\bibitem{sjoberg2012quantifying}
Sj{\o}berg DI, Yamashita A, Anda BC, Mockus A, Dyb{\aa} T. Quantifying the
  effect of code smells on maintenance effort. {\it IEEE Transactions on
  Software Engineering} 2012\string; 39(8)\string: 1144--1156.

\bibitem{neamtiu2013towards}
Neamtiu I, Xie G, Chen J. Towards a better understanding of software evolution:
  an empirical study on open-source software. {\it Journal of Software:
  Evolution and Process} 2013\string; 25(3)\string: 193--218.

\bibitem{chatzigeorgiou2010investigating}
Chatzigeorgiou A, Manakos A. Investigating the evolution of bad smells in
  object-oriented code. {\it 2010 Seventh International Conference on the
  Quality of Information and Communications Technology} 2010\string: 106--115.

\bibitem{menzies2006data}
Menzies T, Greenwald J, Frank A. Data mining static code attributes to learn
  defect predictors. {\it IEEE transactions on software engineering}
  2006\string; 33(1)\string: 2--13.

\bibitem{tufano2017and}
Tufano M, Palomba F, Bavota G, et al. When and why your code starts to smell
  bad (and whether the smells go away). {\it IEEE Transactions on Software
  Engineering} 2017\string; 43(11)\string: 1063--1088.

\bibitem{arcoverde2011understanding}
Arcoverde R, Garcia A, Figueiredo E. Understanding the longevity of code
  smells: preliminary results of an explanatory survey. {\it Proceedings of the
  4th Workshop on Refactoring Tools} 2011\string: 33--36.

\bibitem{digkas2018developers}
Digkas G, Lungu M, Avgeriou P, Chatzigeorgiou A, Ampatzoglou A. How do
  developers fix issues and pay back technical debt in the apache ecosystem?.
  {\it 2018 IEEE 25th International Conference on Software Analysis, Evolution
  and Reengineering (SANER)} 2018\string: 153--163.

\bibitem{lozano2007assessing}
Lozano A, Wermelinger M, Nuseibeh B. Assessing the impact of bad smells using
  historical information. {\it Ninth international workshop on Principles of
  software evolution: in conjunction with the 6th ESEC/FSE joint meeting}
  2007\string: 31--34.

\bibitem{rapu2004using}
Rapu D, Ducasse S, G{\^\i}rba T, Marinescu R. Using history information to
  improve design flaws detection. {\it Eighth European Conference on Software
  Maintenance and Reengineering, 2004. CSMR 2004. Proceedings.} 2004\string:
  223--232.

\bibitem{tufano2015and}
Tufano M, Palomba F, Bavota G, et al. When and why your code starts to smell
  bad. {\it Proceedings of the 37th International Conference on Software
  Engineering-Volume 1} 2015\string: 403--414.

\bibitem{Fowler1999}
Fowler M, Beck K, Brant J, Opdyke W, Roberts D. {\it {Refactoring: Improving
  the Design of Existing Code}}.
\newblock Addison Wesley.
\newblock 1st~ed. 1999

\bibitem{peters2012evaluating}
Peters R, Zaidman A. Evaluating the lifespan of code smells using software
  repository mining. {\it 2012 16th European Conference on Software Maintenance
  and Reengineering} 2012\string: 411--416.

\bibitem{marcilio2019static}
Marcilio D, Bonif{\'a}cio R, Monteiro E, Canedo E, Luz W, Pinto G. Are static
  analysis violations really fixed? a closer look at realistic usage of
  sonarqube. {\it 2019 IEEE-ACM 27th International Conference on Program
  Comprehension (ICPC)} 2019\string: 209--219.

\bibitem{yamashita2012code}
Yamashita A, Moonen L. Do code smells reflect important maintainability
  aspects?. {\it 2012 28th IEEE international conference on software
  maintenance (ICSM)} 2012\string: 306--315.

\bibitem{yamashita2013exploring}
Yamashita A, Moonen L. Exploring the impact of inter-smell relations on
  software maintainability: An empirical study. {\it Proceedings of the 2013
  International Conference on Software Engineering} 2013\string: 682--691.

\bibitem{digkas2017evolution}
Digkas G, Lungu M, Chatzigeorgiou A, Avgeriou P. The evolution of technical
  debt in the apache ecosystem. In: Springer. ; 2017\string: 51--66.

\bibitem{arvanitou2019monitoring}
Arvanitou EM, Ampatzoglou A, Bibi S, Chatzigeorgiou A, Stamelos I. Monitoring
  Technical Debt in an Industrial Setting. {\it Proceedings of the Evaluation
  and Assessment on Software Engineering} 2019\string: 123--132.

\bibitem{zabardast2020}
Zabardast E, Gonzalez-Huerta J, {\v{S}}mite D. Refactoring, bug fixing, and new
  development effect on technical debt: An industrial case study. {\it 2020
  46th Euromicro Conference on Software Engineering and Advanced Applications
  (SEAA)} 2020\string: 376--384.

\bibitem{stol2018abc}
Stol KJ, Fitzgerald B. The ABC of software engineering research. {\it ACM
  Transactions on Software Engineering and Methodology (TOSEM)} 2018\string;
  27(3)\string: 1--51.

\bibitem{saarimaki2019accuracy}
Saarimaki N, Baldassarre MT, Lenarduzzi V, Romano S. On the accuracy of
  sonarqube technical debt remediation time. {\it 2019 45th Euromicro
  Conference on Software Engineering and Advanced Applications (SEAA)}
  2019\string: 317--324.

\bibitem{guaman2017sonarqube}
Guaman D, Sarmiento P, Barba-Guam{\'a}n L, Cabrera P, Enciso L. SonarQube as a
  tool to identify software metrics and technical debt in the source code
  through static analysis. {\it 7th International Workshop on Computer Science
  and Engineering, WCSE} 2017\string: 171--175.

\bibitem{ISO/IEC/IEEE2010}
ISO/IEC/IEEE . {Systems and software engineering ISO/IEC/IEEE 24765 —
  Vocabulary}. tech. rep., ISO; :   2010.

\bibitem{fowler2018refactoring}
Fowler M. {\it Refactoring: improving the design of existing code}.
\newblock Addison-Wesley Professional .
\newblock 2018.

\bibitem{Fowler2006}
Fowler M. {CodeSmell}. https://martinfowler.com/bliki/CodeSmell.html;  2006.

\bibitem{Brown1998}
Brown WJ, Malveau RC, Mowbray TJ, Wiley J. {\it {AntiPatterns: Refactoring
  Software , Architectures, and Projects in Crisis}}.
\newblock No.~4 in 3Wiley .
\newblock 1998.

\bibitem{NIST2012}
NIST . {NIST Special Publication 800-30 Revision 1 - Guide for Conducting Risk
  Assessments}. Tech. Rep. September, ; :   2012

\bibitem{ISO/IEC2018}
ISO/IEC . {IEC 27005:2018 Information technology–security
  techniques–information security risk management}. Tech. Rep.~0, ISO; :
  2018.

\bibitem{miller2011survival}
Miller~Jr RG. {\it Survival analysis}. 66.
\newblock John Wiley \& Sons .
\newblock 2011.

\bibitem{kaplan1958nonparametric}
Kaplan EL, Meier P. Nonparametric estimation from incomplete observations. {\it
  Journal of the American statistical association} 1958\string; 53(282)\string:
  457--481.

\bibitem{leys2013detecting}
Leys C, Ley C, Klein O, Bernard P, Licata L. Detecting outliers: Do not use
  standard deviation around the mean, use absolute deviation around the median.
  {\it Journal of Experimental Social Psychology} 2013\string; 49(4)\string:
  764--766.

\bibitem{Palomba2018}
Palomba F, Bavota G, Penta MD, Fasano F, Oliveto R, Lucia AD. {On the
  diffuseness and the impact on maintainability of code smells: a large scale
  empirical investigation}. {\it Empirical Software Engineering} 2018\string;
  23(3)\string: 1188--1221.
\newblock \href {\doibase 10.1007/s10664-017-9535-z} {doi:
  10.1007/s10664-017-9535-z}

\bibitem{rios2018tertiary}
Rios N, Mendon{\c{c}}a~Neto dMG, Sp{\'\i}nola RO. A tertiary study on technical
  debt: Types, management strategies, research trends, and base information for
  practitioners. {\it Information and Software Technology} 2018\string;
  102\string: 117--145.

\bibitem{li2015systematic}
Li Z, Avgeriou P, Liang P. A systematic mapping study on technical debt and its
  management. {\it Journal of Systems and Software} 2015\string; 101\string:
  193--220.

\bibitem{alves2016identification}
Alves NS, Mendes TS, Mendon{\c{c}}a dMG, Sp{\'\i}nola RO, Shull F, Seaman C.
  Identification and management of technical debt: A systematic mapping study.
  {\it Information and Software Technology} 2016\string; 70\string: 100--121.

\bibitem{Martin2008}
Martin RC. {\it {Clean Code}}.
\newblock Prentice Hall .
\newblock 2008.

\bibitem{nagappan2013diversity}
Nagappan M, Zimmermann T, Bird C. Diversity in software engineering research.
  In: ACM. ; 2013\string: 466--476.

\end{thebibliography}


\begin{thebibliography}{10}

\bibitem{Hirt1974}
Hirt CW, Amsden AA, Cook JL. An arbitrary {L}agrangian-{E}ulerian computing
  method for all flow speeds.  {\it J {C}omput {P}hys. }1974;14(3):227--253.

\bibitem{Liska2010}
Liska R, Shashkov M, Vachal P, Wendroff B. Optimization-based synchronized
  flux-corrected conservative interpolation (remapping) of mass and momentum
  for arbitrary {L}agrangian-{E}ulerian methods.  {\it J {C}omput {P}hys.
  }2010;229(5):1467--1497.

\bibitem{Taylor1937}
Taylor GI, Green AE. Mechanism of the production of small eddies from large
  ones.  {\it P {R}oy {S}oc {L}ond {A} {M}at. }1937;158(895):499--521.
\newblock \url{https://doi.org/10.1098/rspa.1937.0036},
  \url{http://rspa.royalsocietypublishing.org/content/158/895/499}.

\bibitem{Knupp1999}
Knupp PM. Winslow smoothing on two-dimensional unstructured meshes.  {\it Eng
  {C}omput. }1999;15:263--268.

\bibitem{Kamm2000}
Kamm J. {\it Evaluation of the {S}edov-von {N}eumann-{T}aylor blast wave
  solution. } Technical {R}eport LA-UR-00-6055: Los {A}lamos {N}ational
  {L}aboratory; 2000.

\bibitem{Kucharik2003}
Kucharik M, Shashkov M, Wendroff B. An efficient linearity-and-bound-preserving
  remapping method.  {\it J {C}omput {P}hys. }2003;188(2):462--471.

\bibitem{Blanchard2015}
Blanchard G, Loubere R. {\it High-Order {C}onservative {R}emapping with a
  posteriori {MOOD} stabilization on polygonal meshes. }
  \url{https://hal.archives-ouvertes.fr/hal-01207156}, the {HAL} {O}pen
  {A}rchive, hal-01207156. Accessed January 13, 2016; 2015.

\bibitem{Burton2013}
Burton DE, Kenamond MA, Morgan NR, Carney TC, Shashkov MJ. An intersection
  based {ALE} scheme {(xALE)} for cell centered hydrodynamics {(CCH)}.  In:
  Talk at {M}ultimat 2013, {I}nternational {C}onference on {N}umerical
  {M}ethods for {M}ulti-{M}aterial {F}luid {F}lows; September 2--6, 2013; San
  {F}rancisco.
\newblock LA-UR-13-26756.2.

\bibitem{Berndt2011}
Berndt M, Breil J, Galera S, Kucharik M, Maire PH, Shashkov M. Two-step hybrid
  conservative remapping for multimaterial arbitrary {L}agrangian-{E}ulerian
  methods.  {\it J {C}omput {P}hys. }2011;230(17):6664--6687.

\bibitem{Kucharik2012}
Kucharik M, Shashkov M. One-step hybrid remapping algorithm for multi-material
  arbitrary {L}agrangian-{E}ulerian methods.  {\it J {C}omput {P}hys.
  }2012;231(7):2851--2864.

\bibitem{Breil2015}
Breil J, Alcin H, Maire PH. A swept intersection-based remapping method for
  axisymmetric {ReALE} computation.  {\it Int {J} {N}umer {M}eth {F}l.
  }2015;77(11):694--706.
\newblock Fld.3996.

\bibitem{Barth1997}
Barth TJ. Numerical methods for gasdynamic systems on unstructured meshes.  In:
   Kroner D, Rohde C, Ohlberger M, eds. {\it An {I}ntroduction to {R}ecent
  {D}evelopments in {T}heory and {N}umerics for {C}onservation {L}aws,
  {P}roceedings of the {I}nternational {S}chool on {T}heory and {N}umerics for
  {C}onservation {L}aws}, Lecture {N}otes in {C}omputational {S}cience and
  {E}ngineering. Berlin: Springer 1997.
\newblock ISBN 3-540-65081-4.

\bibitem{Lauritzen2011}
Lauritzen P, Erath C, Mittal R. On simplifying `incremental remap'-based
  transport schemes.  {\it J {C}omput {P}hys. }2011;230(22):7957--7963.

\bibitem{Klima2017}
Klima M, Kucharik M, Shashkov M. Local error analysis and comparison of the
  swept- and intersection-based remapping methods.  {\it Commun {C}omput
  {P}hys. }2017;21(2):526--558.

\bibitem{Dukowicz2000}
Dukowicz JK, Baumgardner JR. Incremental remapping as a transport/advection
  algorithm.  {\it J {C}omput {P}hys. }2000;160(1):318--335.

\bibitem{Kucharik2011}
Kucharik M, Shashkov M. Flux-based approach for conservative remap of
  multi-material quantities in {2D} arbitrary {L}agrangian-{E}ulerian
  simulations.  In:  Fo\v{r}t J, F{\"{u}}rst J, Halama J, Herbin R, Hubert F,
  eds. {\it Finite {V}olumes for {C}omplex {A}pplications {VI} {P}roblems \&
  {P}erspectives},  Springer {P}roceedings in {M}athematics, vol. 1: Springer
  2011 (pp. 623--631).

\bibitem{Kucharik2014}
Kucharik M, Shashkov M. Conservative multi-material remap for staggered
  multi-material arbitrary {L}agrangian-{E}ulerian methods.  {\it J {C}omput
  {P}hys. }2014;258:268--304.

\bibitem{Loubere2005}
Loubere R, Shashkov M. A subcell remapping method on staggered polygonal grids
  for arbitrary-{L}agrangian-{E}ulerian methods.  {\it J {C}omput {P}hys.
  }2005;209(1):105--138.

\bibitem{Caramana1998}
Caramana EJ, Shashkov MJ. Elimination of artificial grid distortion and
  hourglass-type motions by means of {L}agrangian subzonal masses and
  pressures.  {\it J {C}omput {P}hys. }1998;142(2):521--561.

\bibitem{Hoch2009}
Hoch P. {\it An arbitrary {L}agrangian-{E}ulerian strategy to solve
  compressible fluid flows. } Technical {R}eport: CEA; 2009.
\newblock HAL: hal-00366858.
  https://hal.archives-ouvertes.fr/docs/00/36/68/58/PDF/ale2d.pdf. Accessed
  January 13, 2016.

\bibitem{Shashkov1996}
Shashkov M. {\it Conservative {F}inite-{D}ifference {M}ethods on {G}eneral
  {G}rids}.
\newblock Boca Raton, Florida: CRC {P}ress; 1996.
\newblock ISBN 0-8493-7375-1.

\bibitem{Benson1992}
Benson DJ. Computational methods in {L}agrangian and {E}ulerian hydrocodes.
  {\it Comput {M}ethod {A}ppl {M}. }1992;99(2--3):235--394.

\bibitem{Margolin2003}
Margolin LG, Shashkov M. Second-order sign-preserving conservative
  interpolation (remapping) on general grids.  {\it J {C}omput {P}hys.
  }2003;184(1):266--298.

\bibitem{Kenamond2013}
Kenamond MA, Burton DE. Exact intersection remapping of multi-material
  domain-decomposed polygonal meshes.  In: Talk at {M}ultimat 2013,
  {I}nternational {C}onference on {N}umerical {M}ethods for {M}ulti-{M}aterial
  {F}luid {F}lows; September 2--6, 2013; San {F}rancisco.
\newblock LA-UR-13-26794.

\bibitem{Dukowicz1984}
Dukowicz J. Conservative rezoning (remapping) for general quadrilateral meshes.
   {\it J {C}omput {P}hys. }1984;54(3):411--424.

\bibitem{Margolin2002}
Margolin LG, Shashkov M. {\it Second-order sign-preserving remapping on general
  grids. } Technical Report LA-UR-02-525: Los {A}lamos {N}ational {L}aboratory;
  2002.

\bibitem{Mavriplis2003}
Mavriplis DJ. Revisiting the least-squares procedure for gradient
  reconstruction on unstructured meshes.  In: AIAA 2003-3986. 16th {AIAA}
  {C}omputational {F}luid {D}ynamics {C}onference; June 23--26, 2003; Orlando,
  {F}lorida.

\bibitem{Scovazzi2008}
Scovazzi G, Love E, Shashkov M. Multi-scale {L}agrangian shock hydrodynamics on
  {Q1/P0} finite elements: {T}heoretical framework and two-dimensional
  computations.  {\it Comput {M}ethod {A}ppl {M}. }2008;197(9--12):1056--1079.

\end{thebibliography}
